\documentclass[lettersize,journal]{IEEEtran}
\usepackage{amsmath,amsfonts}
\usepackage{algorithmic}
\usepackage{algorithm}
\usepackage{array}
\usepackage[caption=false,font=normalsize,labelfont=sf,textfont=sf]{subfig}
\usepackage{textcomp}
\usepackage{stfloats}
\usepackage{url}
\usepackage{verbatim}
\usepackage{graphicx}
\usepackage{cite}
\usepackage{color}
\usepackage{booktabs}
\usepackage{hyperref}
\def\BibTeX{{\rm B\kern-.05em{\sc i\kern-.025em b}\kern-.08em
    T\kern-.1667em\lower.7ex\hbox{E}\kern-.125emX}}
\usepackage{balance}

\begin{document}
\title{Experimental Study of RCS Diversity with Novel No-divergent OAM Beams}
\author{Yufei Zhao,
        Yong Liang Guan,~\IEEEmembership{Senior Member,~IEEE,}
        Dong Chen,
        Afkar Mohamed Ismail,
        Xiaoyan Ma,~\IEEEmembership{Member,~IEEE,}
        Xiaobei Liu,~\IEEEmembership{Member,~IEEE,}
        and~Chau Yuen,~\IEEEmembership{Fellow,~IEEE}
\thanks{This work has been submitted to IEEE Transactions on Vehicular Technology. Manuscript received, Jul., 2024; revised Oct, ** 2024. This work was supported by the National Research Foundation, Singapore, and Infocomm Media Development Authority under its Future Communications Research \& Development Programme, Grant No. FCP-NTU-RG-2022-011 and Grant No. FCP-NTU-RG-2022-020. (Corresponding author: Yong Liang Guan).}
\thanks{Yufei Zhao, Yong Liang Guan, Afkar Mohamed Ismail, Xiaoyan Ma and Chau Yuen are with the School of Electrical and Electronic Engineering, Nanyang Technological University, 639798, Singapore (e-mail: \{yufei.zhao, eylguan, afkar.mi, xiaoyan.ma, chau.yuen\}@ntu.edu.sg).

Dong Chen is with the School of Aerospace Engineering, Tsinghua University, Beijing, China (email: dong\_chen\_2126@yeah.net).

Xiaobei Liu is with the Temasek Laboratories@NTU, Nanyang Technological University, 639798, Singapore (email: xpliu@ntu.edu.sg).

}
}
\markboth{IEEE Transactions on Vehicular Technology,~Vol.~**, No.~**, **~2024}%
{Yufei Zhao \MakeLowercase{\textit{et al.}}: A Sample Article Using IEEEtran.cls for IEEE Journals}




\maketitle

\begin{abstract}
This research proposes a novel approach utilizing Orbital Angular Momentum (OAM) beams to enhance Radar Cross Section (RCS) diversity for target detection in future transportation systems. Unlike conventional OAM beams with hollow-shaped divergence patterns, the new proposed OAM beams provide uniform illumination across the target without a central energy void, but keep the inherent phase gradient of vortex property. We utilize waveguide slot antennas to generate four different modes of these novel OAM beams at X-band frequency. Furthermore, these different mode OAM beams are used to illuminate metal models, and the resulting RCS is compared with that obtained using plane waves. The findings reveal that the novel OAM beams produce significant azimuthal RCS diversity, providing a new approach for the detection of weak and small targets.
This study not only reveals the RCS diversity phenomenon based on novel OAM beams of different modes but also addresses the issue of energy divergence that hinders traditional OAM beams in long-range detection applications.
\end{abstract}


\section{Introduction}
%
%
%
%
Radar Cross Section (RCS) represents the target's ability to reflect radar signals, which depends on various factors, including the target's size, shape, material, and orientation relative to the radar. In a radar system, the RCS helps determine how easily a target can be detected and identified. Traditional radar systems utilize plane waves to illuminate targets, and the RCS is calculated based on the backscattered signal received by the radar. However, even for the same target, the backscattering characteristics can vary significantly with different angles of incidence and observation. This variability makes the study of RCS diversity crucial for improving the detection and identification of various targets in a three-dimensional traffic network \cite{TVT2}. Given that roadside Road-Side Units (RSUs) are typically fixed and non-mobile, they can only illuminate targets from a specific, fixed angle. This limitation poses a challenge for achieving RCS diversity using conventional plane waves. Without RCS diversity, the RSUs might struggle to distinguish between different targets or to detect small targets effectively, e.g., drones,  which can undermine the reliability of the intelligent transportation system \cite{Air}.

Orbital Angular Momentum (OAM) waves have emerged as a promising solution to address these challenges. OAM Electro-Magnetic (EM) waves are characterized by their helical wavefronts, which create a phase gradient perpendicular to the propagation direction \cite{TWC,Jie}. This unique property allows OAM waves to introduce new dimensions in target detection and identification. Unlike plane waves, the helical structure of OAM waves provides multiple modes, each with a distinct wavefront distribution, potentially enhancing RCS diversity. In 2017, Zhang \textit{et al.} explored the RCS characteristics of OAM waves and highlighted their potential for anti-stealth radar applications by showing that different modes can significantly affect backscatter patterns \cite{Dong}.
In, 2020, Liu \textit{et al.} investigated the backward scattering of OAM beams by electrically large objects and found that the scattered fields retain the vortex characteristics of the incident OAM \cite{Liu}.

Although there has been extensive research on target echo detection using OAM waves, most studies have focused on theoretical modeling and experimental analysis based on traditional hollow-shaped vortex beams, with energy primarily distributed in a ring shape \cite{AWPL1,AWPL2}.
As the distance increases, the energy dispersion becomes significant, which is a major limitation for the long-distance transmission of OAM beams, especially in radar applications where long-range detection is crucial \cite{Chen1}. To address this tricky issue, this paper proposes a novel OAM beam without a central void and applies it to radar target detection.
The enhanced RCS diversity comes from the wavefront changes of various OAM modes, which is crucial for accurately detecting and identifying various traffic units, thus ensuring better coordination and management of data transmission in intelligent transportation systems. Furthermore, this study involves a combination of experimental measurements and full-wave simulations. The experimental setup includes generating OAM beams with different topology modes using the waveguide slot antennas, and measurement RCS of various targets at X-band in an anechoic chamber.

\section{Novel OAM Beams without Energy Void}
\begin{figure}[htbp]
\centering
\includegraphics[width=3.4in]{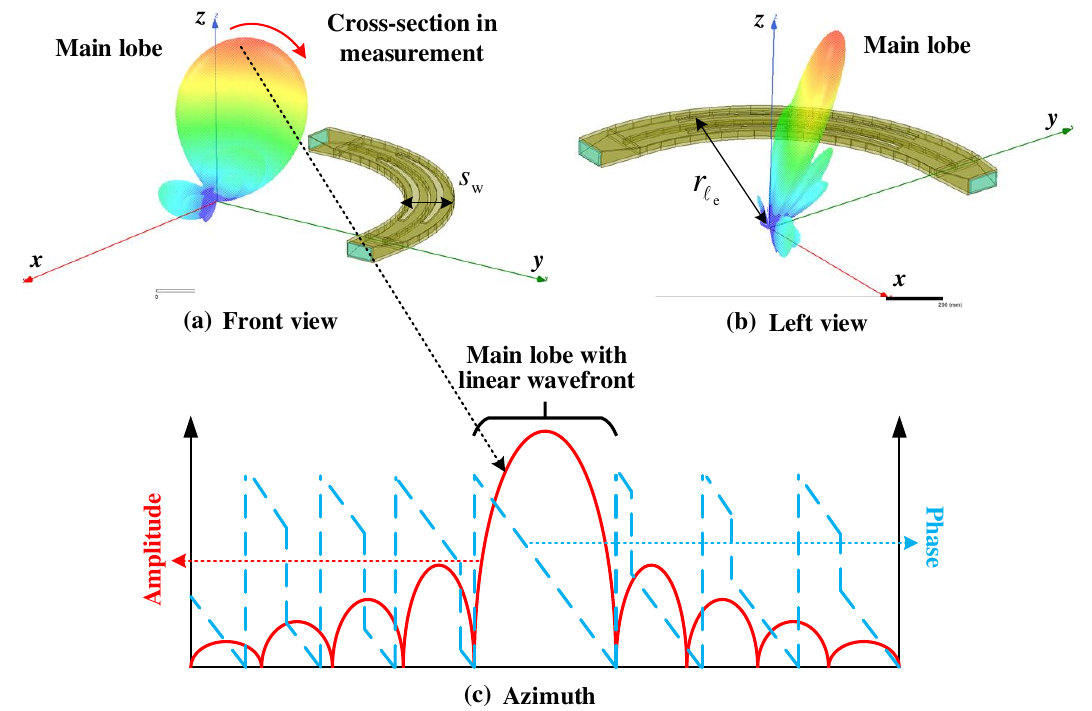}
\caption{A NTCS-OAM generator design without central energy void with respect to (a) the 3D radiation pattern from the front view, (b) the same radiation pattern from the left view (c) 2D beam schematic diagram showing the amplitude (solid line) and the phase (dashed line) distributions.}
\label{wavefront}
\end{figure}

In the spatial domain, as we know, the OAM waves exhibit an azimuthal phase distribution of $\exp \left( { - {\rm{j}}\ell \varphi } \right)$, and different OAM modes $\ell$ introduce periodic spatial phase variations \cite{Wenchi1}. By superimposing these modes, we can achieve a far-field radiation pattern with arbitrary angular intensity and phase distributions \cite{Xianmin}. Thus, this method provides a directional gain in the constructed radio beams while maintaining the phase variations associated with OAM modes in the radiation direction. This offers theoretical guidance for generating high-gain OAM pencil beams.
By leveraging this property, recently, a Non-Uniform Traveling-Wave Current Sources (NTCS) method has been proposed to generate directional OAM beams, which is equivalent to superimposing different OAM modes together, thereby achieving a far-field radiation pattern with arbitrary angular intensity and phase distributions \cite{zhu}. Inspired by this principle, our research proposes a novel practical leaky-wave waveguide antenna, which can generate high-gain OAM waves with arbitrary modes, featuring a directional pattern with high directivity and concentrated energy, as shown in Fig. \ref{wavefront}. This new type of generator retains the wavefront phase variation of OAM beams but lacks the hollow-shaped radiation pattern.

As depicted in Fig. \ref{wavefront}(a)(b), the waveguide generator has a partially circular arc shape. One end of the waveguide is excited by a signal source while the other end is terminated with a matching load, establishing a traveling-wave environment. The waveguide comprises a wide side ${s_{\rm{w}}}$ and a narrow side ${s_{\rm{n}}}$, operating under the ${\rm{TE}}_{10}$ mode. By creating slots along the wide side, a slotted leaky-wave antenna is constructed. The relationship between the cutoff wavelength $\lambda_{\rm{c}}^{\rm{TE}_{10}}$ and the cavity dimensions is
$\lambda _{\rm{c}}^{{\rm{T}}{{\rm{E}}_{{\rm{10}}}}} = {\lambda _0}/\sqrt {1 - {{\left( {{\lambda _0}/2{s_{\rm{w}}}} \right)}^2}}$,
where $\lambda_0$ is the wavelength in vacuum. According to the definition of OAM topological mode, we know that the equivalent mode number can be denoted as $\ell  = 2\pi a/{\lambda _g}$. From the perspective of actual leaky-wave cavity design, $a$ is the radius that equals to ${r_{{l_e}}} + {s_w}/2$ shown in Fig. \ref{wavefront}, ${\lambda _g}$ is the guided wavelength that is equal to $\lambda _{\rm{c}}^{{\rm{T}}{{\rm{E}}_{{\rm{10}}}}}$. Hence, there is a relationship between the physical dimensions and equivalent OAM mode ${\ell_{\rm{e}}}$ as
\begin{equation} \label{new_eq2}
\begin{array}{l}
\left| {{l_{\rm{e}}}} \right| = \frac{{2\pi a}}{{{\lambda _g}}} = \frac{{2\pi \left( {{r_{{l_e}}} + {{{s_w}} \mathord{\left/
 {\vphantom {{{s_w}} 2}} \right.
 \kern-\nulldelimiterspace} 2}} \right)}}{{\lambda _{\rm{c}}^{{\rm{T}}{{\rm{E}}_{{\rm{10}}}}}}}\\
{\kern 1pt} {\kern 1pt} {\kern 1pt} {\kern 1pt} {\kern 1pt} {\kern 1pt} {\kern 1pt} {\kern 1pt} {\kern 1pt} {\kern 1pt}  = \frac{{2\pi \left( {{r_{{l_e}}} + {{{s_w}} \mathord{\left/
 {\vphantom {{{s_w}} 2}} \right.
 \kern-\nulldelimiterspace} 2}} \right)}}{{{{{\lambda _0}} \mathord{\left/
 {\vphantom {{{\lambda _0}} {\sqrt {1 - {{\left( {\frac{{{\lambda _0}}}{{2{s_w}}}} \right)}^2}} }}} \right.
 \kern-\nulldelimiterspace} {\sqrt {1 - {{\left( {\frac{{{\lambda _0}}}{{2{s_w}}}} \right)}^2}} }}}} = \frac{{2\pi \left( {{r_{{l_e}}} + {{{s_w}} \mathord{\left/
 {\vphantom {{{s_w}} 2}} \right.
 \kern-\nulldelimiterspace} 2}} \right)\sqrt {1 - {{\left( {\frac{{{\lambda _0}}}{{2{s_w}}}} \right)}^2}} }}{{{\lambda _0}}}\\
{\kern 1pt} {\kern 1pt} {\kern 1pt} {\kern 1pt} {\kern 1pt} {\kern 1pt} {\kern 1pt} {\kern 1pt} {\kern 1pt} {\kern 1pt}  = \left( {{r_{{l_e}}} + {{{s_w}} \mathord{\left/
 {\vphantom {{{s_w}} 2}} \right.
 \kern-\nulldelimiterspace} 2}} \right)\pi \sqrt {{{\left( {\frac{2}{{{\lambda _0}}}} \right)}^2} - {{\left( {\frac{1}{{{s_w}}}} \right)}^2}}
\end{array},
\end{equation}
which will guide us in designing these NTCS-OAM antennas with specific dimensions based on the desired OAM mode ${\ell_{\rm{e}}}$. As illustrated in Fig. \ref{wavefront}(c), these new OAM beam exhibits a linear phase variation within the main lobe, determined by their topological mode number, which is its most noticeable difference compared to plane waves.
Meanwhile, like a plane wave antenna, the OAM beam generated by our novel NTCS-OAM antenna also has a distinct main radiation lobe, enhancing its practical viability for long-range radar applications.
It should be noted that there are various configurations of waveguide leaky-wave antennas that can generate directional radiated OAM beams \cite{Xianmin}. In this paper, we only utilize a kind of $90^\circ$ arc-shaped structure as a practical example.

\begin{figure*}[htbp]
\centering
\includegraphics[width=7.2in]{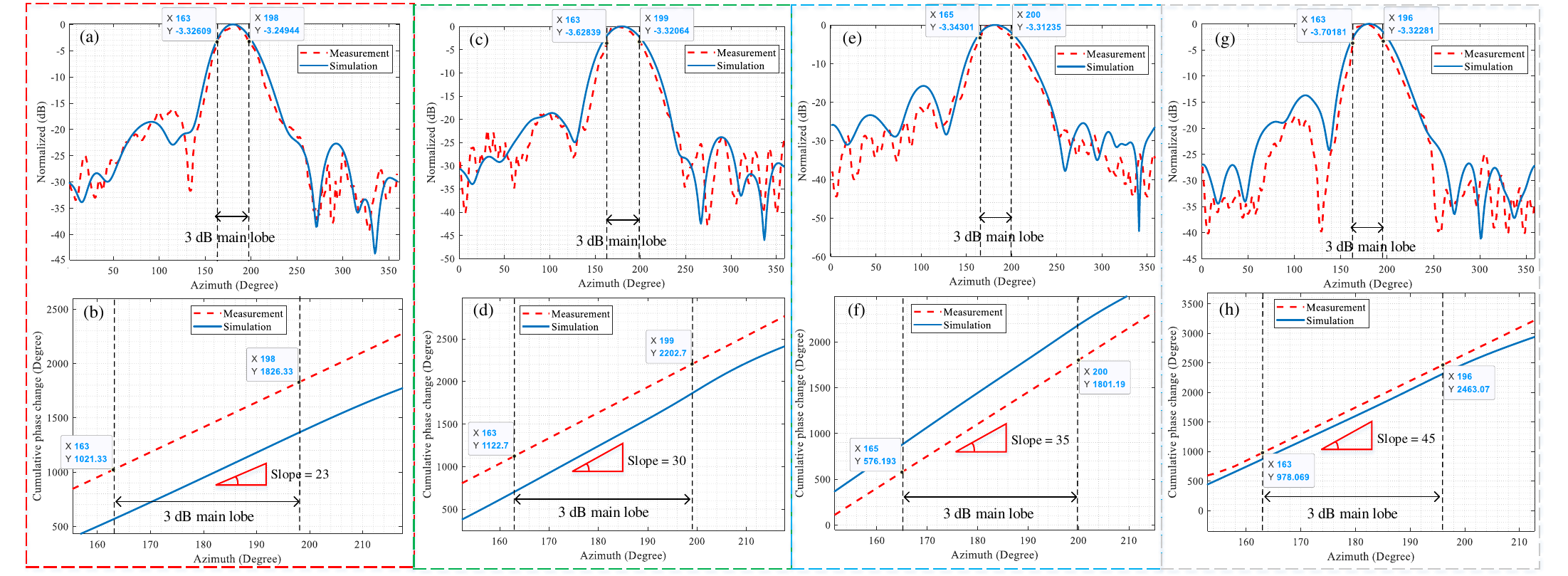}
\caption{Novel 2D radiation pattern of OAM mode with respect to (a) ${\ell _{\rm{e}}} = 23$, (b) ${\ell _{\rm{e}}} = 30$, (c) ${\ell _{\rm{e}}} = 35$, (d) ${\ell _{\rm{e}}} = 45$, comparing results between simulation (solid line) and measurement (dashed line).}
\label{OAM}
\end{figure*}

The simulation process is carried out using Ansys HFSS software. Then, we fabricate the real antenna model sequentially mounted each OAM transmitter, working at 10 GHz, of varying modes onto a turntable in the microwave anechoic chamber. Following simulation guidelines, we tilted the partial arc NTCS-OAM antenna by approximately ${18^ \circ }$ and allowed it to rotate ${360^ \circ }$ around its geometric center. As shown in Fig. \ref{OAM}, both the amplitude envelopes and phase distributions within the 3 dB main lobe of the measurement results closely match the simulation curves. Clearly, within the 3dB beamwidth of the main lobe, the wavefront phase varies linearly, and the slope of the measured curve aligns well with the simulation results. The slope of the wavefront phase change is determined by the equivalent OAM topological mode number. For instance, from the measurements in Fig. \ref{OAM}(a), we can calculate that the equivalent OAM topological mode number of the directed OAM beam radiated by the NTCS antenna is ${\ell _{\rm{e}}} = \left( {{{1826}^ \circ } - {{1021}^ \circ }} \right)/\left( {{{198}^ \circ } - {{163}^ \circ }} \right) = 23$. The calculation methods for other modes of OAM antennas are similar, indicating that the OAM beams exhibit different phase changes within the main lobe of directional radiation patterns.

It should be noted that the NTCS-OAM antenna proposed in this paper is merely an example. In practice, numerous antenna design methods can be used to generate this pencil-shaped OAM beam, including discrete array antennas \cite{TAP}, dielectric resonator antennas, and metasurfaces \cite{IoT1}, etc. The fundamental guiding principle behind this design is a novel wavefront manipulation technique, which is a key focus of this work. In traditional antenna design, or in communication and sensing applications, attention is typically given to the shape of the radiation pattern, optimizing factors such as gain, bandwidth, sidelobes, and axial ratio. However, in addition to these considerations, there also exists phase variation within the main lobe of the antenna radiation. As demonstrated in this study, by adopting specific design approaches, it is possible to flexibly control the phase variation of the wavefront within the main lobe of the radiation \cite{IoT2}. This introduces a new degree of freedom for physical manipulation, which holds significant potential for future wireless communication and sensing applications. We are currently continuing our research in this area to further explore these possibilities.

\section{Modeling for ``Simple'' Complicated Targets}
\begin{figure}[htbp]
\centering
\includegraphics[width=3.0in]{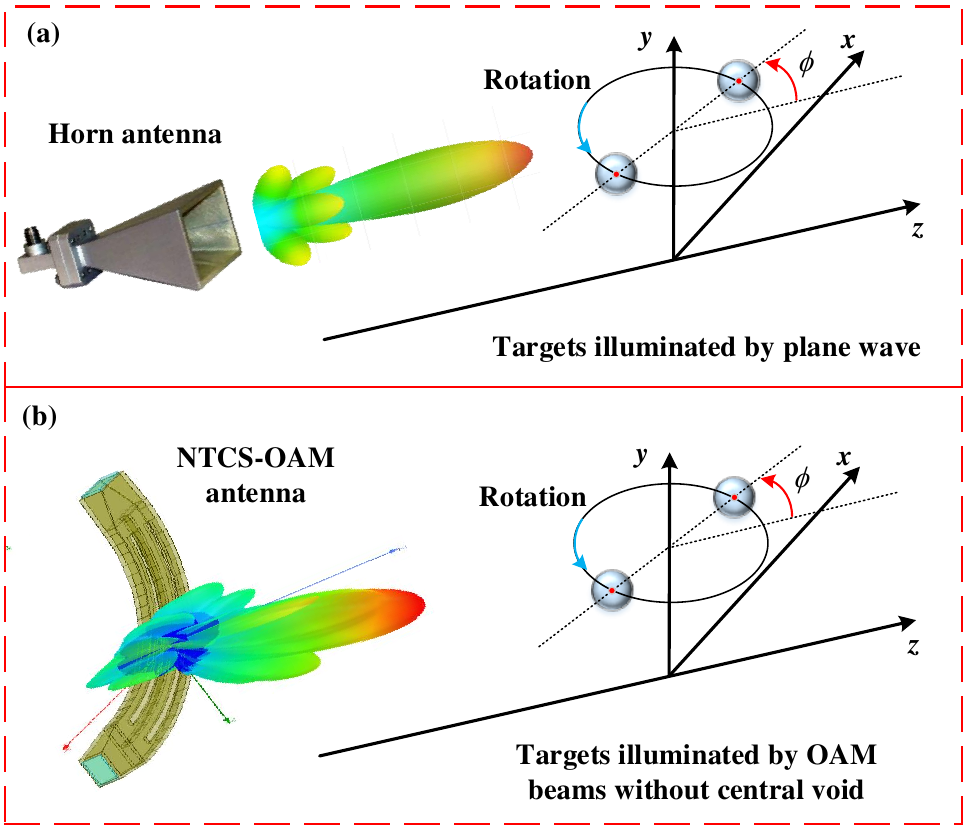}
\caption{The EM wave illuminates a ``simple'' complex target consisting of two identical metal spheres, with respect to (a) by the conventional plane wave, and (b) by the novel OAM beam without central void.}
\label{plane_OAM}
\end{figure}

As we know, a target whose echo is within a range gate is called a point target. Point targets are further divided into simple targets and complex targets. A simple target contains only a single scatterer, while a complex target consists of multiple scatterers \cite{Dong,book}.
The traditional RCS fluctuations of plane waves consider only changes in radar viewing angle and electromagnetic wave frequency. As shown in Fig. \ref{plane_OAM}(a), under these circumstances, a ``simple'' complex target composed of two identical scatterers (such as metal spheres) satisfies the relationship between the RCS fluctuations and the observation angle $\phi $ under plane wave illumination \cite{book}
\begin{equation} \label{eq3}
\frac{\sigma_r}{\sigma_0} = 2 \left[ 1 + \cos(2kD \cos \phi) \right],
\end{equation}
where $\sigma_0$ is the RCS of a single metal sphere, $\sigma_r$ is the RCS of the entire scatterer, $k$ is the wave vector, and $D$ is the distance between the two scatterers. If the plane wave is replaced with an OAM beam, the scenario is shown in Fig. \ref{plane_OAM}(b), illustrating the relative position relation of the targets under the illumination of the OAM main lobe. The distance $D$ is the spacing between the two small spheres, which are at the $x-y$ plane, and $\phi$ is the angle between the line connecting the two small spheres and the $z$-axis.

Suppose that there are $N$ scattering points that form a ``simple'' complex target, each isotropic scatterer in a Cartesian coordinate system with coordinates $[x_n, y_n, z_n]$. In the scenario shown in Fig. \ref{plane_OAM}, they are specifically as
\begin{equation} \label{eq4}
\begin{footnotesize}
\begin{gathered}
{P_n} = {\left[ {\frac{D}{2}\sin {\phi _n},{\mkern 1mu} {y_0},{\mkern 1mu} \frac{D}{2}\cos {\phi _n}} \right]^{\rm{T}}},{\phi _n} = \arctan \left( {\frac{{{y_k}}}{{{x_k}}}} \right).
\end{gathered}
\end{footnotesize}
\end{equation}
Assuming the incident wave is an OAM wave, since each scatterer is isotropic, the reflected wave can be considered a spherical wave. The RCS of each scatterer after being modulated by the amplitude and phase of the OAM EM wave, then the total echo signal can be calculated as
\begin{equation} \label{eq5}
\begin{footnotesize}
\begin{gathered}
{E_s} = \frac{1}{{2\sqrt {\pi R} }}\sqrt {{\sigma _0}} \sum\limits_{n = 1}^N {{g_\ell }\left( {{\theta _n},{\varphi _n}} \right){e^{{\rm{j}}\ell {\rm{arctan}}\left( {\frac{{{y_n}}}{{{x_n}}}} \right)}}{e^{{\rm{j}}{k_z}{z_n}}}{e^{{\rm{j}}k{z_n}}}} ,
\end{gathered}
\end{footnotesize}
\end{equation}
where, $k_z$ is the component of the wave vector $k$ in the $z$-axis direction, ${{g_\ell }\left( {{\theta _n},{\varphi _n}} \right)}$ denotes the radiation gain of the novel OAM beam with mode number $\ell $, ${\theta _n}  = {\rm{arctan}}\left( {{y_n}/{x_n}} \right)$, ${\varphi _n}  = \arctan \left( {\sqrt {x_n^2 + y_n^2} /z_n^2} \right)$, both ${{g_\ell }\left( {{\theta _n},{\varphi _n}} \right)}$ and $\ell$ can be measured by the full wave simulation through HFSS or CST. For the incident electric field density $E_i$, it's mean amplitude can be denoted as
\begin{equation} \label{eq6}
\begin{footnotesize}
\begin{gathered}
\left| {\overline {{E_i}} } \right| = \frac{1}{{2\sqrt {\pi R} }}\sqrt {{\sigma _0}} \frac{1}{N}\sum\limits_{n = 1}^N {{g_\ell }\left( {{\theta _n},{\varphi _n}} \right)} ,
\end{gathered}
\end{footnotesize}
\end{equation}
hence, the RCS scale of the OAM EM wave with mode number $\ell$ can be denoted as
\begin{equation} \label{eq7}
\begin{footnotesize}
\begin{gathered}
\frac{{{\sigma _\ell }}}{{{\sigma _0}}} = \frac{{4\pi {R^2}}}{{{\sigma _0}}}{\left| {\frac{{{E_s}}}{{{E_i}}}} \right|^2} = {N^2}\frac{{{{\left\| {\sum\limits_{n = 1}^N {{g_n}\left( {\theta ,\varphi } \right)} {e^{{\rm{j}}\ell {\rm{arctan}}\left( {\frac{{{y_n}}}{{{x_n}}}} \right)}}{e^{{\rm{j}}{k_z}{z_n}}}{e^{{\rm{j}}k{z_n}}}} \right\|}^2}}}{{{{\left\| {\sum\limits_{n = 1}^N {{g_\ell }\left( {{\theta _n},{\varphi _n}} \right)} } \right\|}^2}}}.
\end{gathered}
\end{footnotesize}
\end{equation}
When simplified to the dual-sphere model shown in Fig. \ref{plane_OAM}, the specific value is $K = 2$. The coordinates of sphere no. 1 and no. 2 are
\begin{equation} \label{eq8}
\begin{footnotesize}
\begin{gathered}
{\left[ { - \frac{D}{2}\sin \phi ,{y_0},\frac{D}{2}\cos \phi } \right]^{\rm{T}}},{\kern 1pt} {\left[ {\frac{D}{2}\sin \phi ,{y_0}, - \frac{D}{2}\cos \phi } \right]^{\rm{T}}}.
\end{gathered}
\end{footnotesize}
\end{equation}
Substituting the coordinates of the two spheres into equation \eqref{eq7}, and eliminating the gain functions from the numerator and denominator, we can get
\begin{equation} \label{eq9}
\frac{{{\sigma _{\ell} }}}{{{\sigma _0}}} \approx {\left| \begin{array}{l}
{e^{ - {\rm{j}}\ell \arctan \left( {\frac{{D\sin \phi }}{{2{y_0}}}} \right)}}{e^{{\rm{j}}{k_z}\frac{D}{2}\cos \phi }}{e^{{\rm{j}}k\frac{D}{2}\cos \phi }}\\
 + {e^{{\rm{j}}\ell \arctan \left( {\frac{{D\sin \phi }}{{2{y_0}}}} \right)}}{e^{ - {\rm{j}}{k_z}\frac{D}{2}\cos \phi }}{e^{ - {\rm{j}}k\frac{D}{2}\cos \phi }}
\end{array} \right|^2},
\end{equation}
simplified, and in the case of $k_z \approx k$, it becomes
\begin{equation} \label{eq10}
\begin{footnotesize}
\begin{gathered}
\frac{{{\sigma _{\ell}}}}{{{\sigma _0}}} \approx 2\left[ {1 + \cos \left( {2kD\cos \phi  - 2\ell \arctan \left( {\frac{{D\sin \phi }}{{2{y_0}}}} \right)} \right)} \right].
\end{gathered}
\end{footnotesize}
\end{equation}

From equation \eqref{eq10}, it can be seen that, unlike plane waves, the RCS of a target under OAM EM wave illumination is related to the mode number $\ell $ and the spatial position. When each sub-scatterer is not isotropic, i.e., the RCS of each sub-scatterer is a function of the orientation angle $\phi$, denoted as ${\sigma _n}\left( \phi  \right)$, the RCS can be expressed as
\begin{equation} \label{eq11}
\begin{footnotesize}
\begin{gathered}
{\sigma _\ell }\left( \phi  \right) = {N^2}{\left| {\frac{{\sum\limits_{n = 1}^N {\sqrt {{\sigma _n}\left( \phi  \right)} } {g_\ell }\left( {{\theta _n},{\varphi _n}} \right){e^{{\rm{j}}\ell \arctan \left( {\frac{{{y_n}}}{{{x_n}}}} \right)}}{e^{j{k_z}{z_n}}}{e^{jk{z_n}}}}}{{\sum\limits_{n = 1}^N {{g_\ell }\left( {{\theta _n},{\varphi _n}} \right)} }}} \right|^2}.
\end{gathered}
\end{footnotesize}
\end{equation}

\begin{figure}[htbp]
\centering
\includegraphics[width=3.5in]{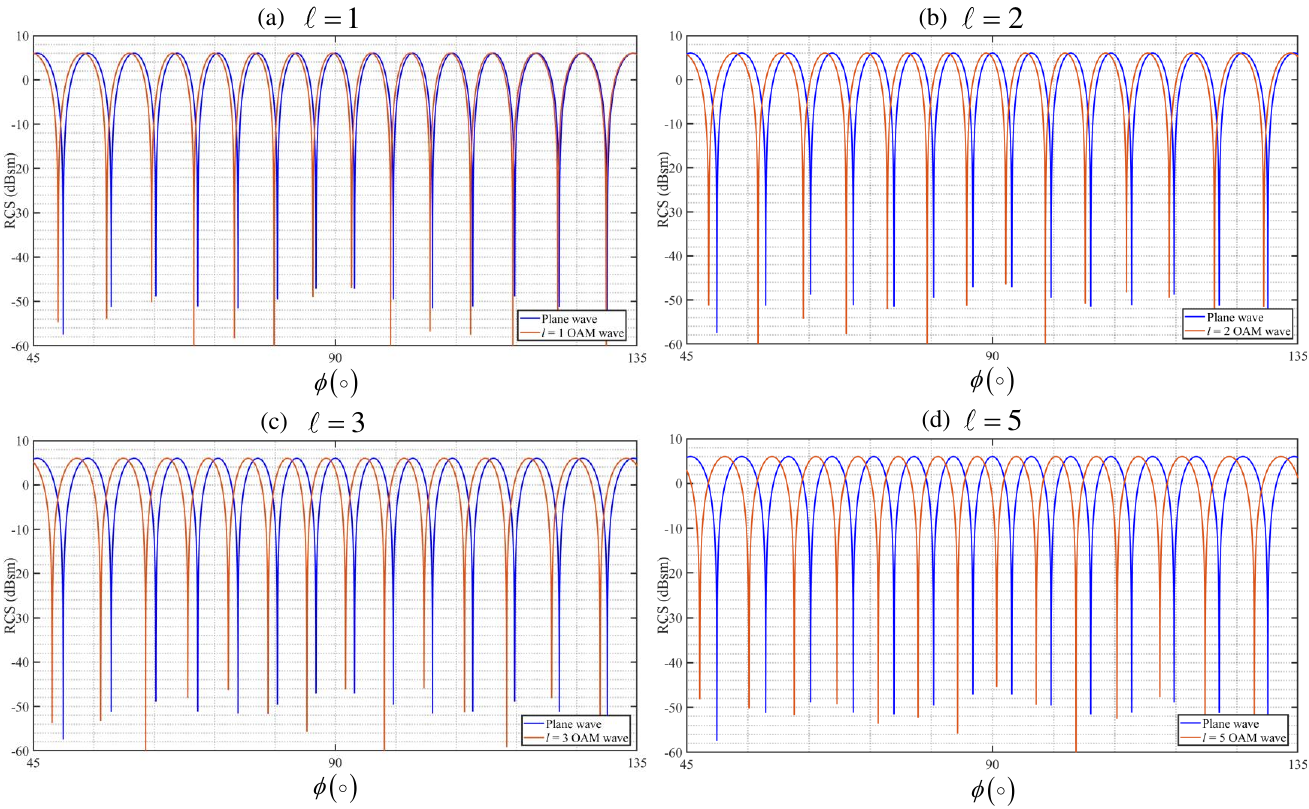}
\caption{RCS results obtained when OAM beams with different mode numbers illuminate a ``simple'' complex target consisting of two identical metal spheres, which are compared to those obtained with plane waves. (a) Comparing with OAM mode $\ell  = 1$. (b) Comparing with OAM mode $\ell  = 2$. (c) Comparing with OAM mode $\ell  = 3$. (d) Comparing with OAM mode $\ell  = 5$.}
\label{OAM_simulation}
\end{figure}

The above theoretically proves that the RCS of OAM EM waves is different from that of plane waves. It can be seen from \eqref{eq10} that the RCS of OAM EM waves is influenced not only by the term $\alpha \left( \phi  \right) = 2kD\cos \phi$ but also by
$\beta \left( \phi  \right) = 2\ell \arctan \left( {\frac{{D\sin \phi }}{{2{y_0}}}} \right)$,
and both $\alpha \left( \phi  \right)$ and $\beta \left( \phi  \right)$ are periodic functions with a period of $2\pi$. It can be seen that the RCS fluctuation characteristics of OAM EM waves are influenced by two factors, i.e., one is $\alpha \left( \phi  \right)$ due to the radial size of the target, and the other is $\beta \left( \phi  \right)$ due to the lateral size of the target. As the mode number $\ell$ increases, the influence of the lateral size of the target becomes more significant.
To more intuitively illustrate the effect of the OAM mode number on RCS, Fig. \ref{OAM_simulation} shows the one-dimensional RCS curves related only to $\phi $. It can be seen that when $\ell  = 1$, the effect of $\beta \left( \phi  \right)$ is much smaller than $\alpha \left( \phi  \right)$, so the RCS fluctuation characteristics are almost the same as those of plane waves. However, as $\ell$ gradually increases, the difference in fluctuation characteristics between OAM EM waves and plane waves becomes more and more pronounced.

Although the ``simple'' complex target composed of two small spheres is very basic in form, it is an important example for explaining the RCS diversity characteristics of OAM EM waves with various modes. From it, we can obtain an intuitive understanding of the fluctuation characteristics of RCS from different observation directions with OAM waves. Additionally, during applications, employing the NTCS transmitter proposed in this paper is an effective approach for generating multiple OAM modes. By appropriately adjusting the radius of the arc-shaped waveguide, the waveguide width, and the position of the feed point, different modes of pencil-shaped OAM beams can be flexibly generated at the same frequency, as illustrated in Fig. \ref{OAM}. Moreover, the gain of the main lobe and radiation directions of these different pencil-shaped OAM beams remain largely consistent, which facilitates various experimental validations.

\section{Experiment Setups and Results}
\begin{figure}[htbp]
\centering
\includegraphics[width=3.4in]{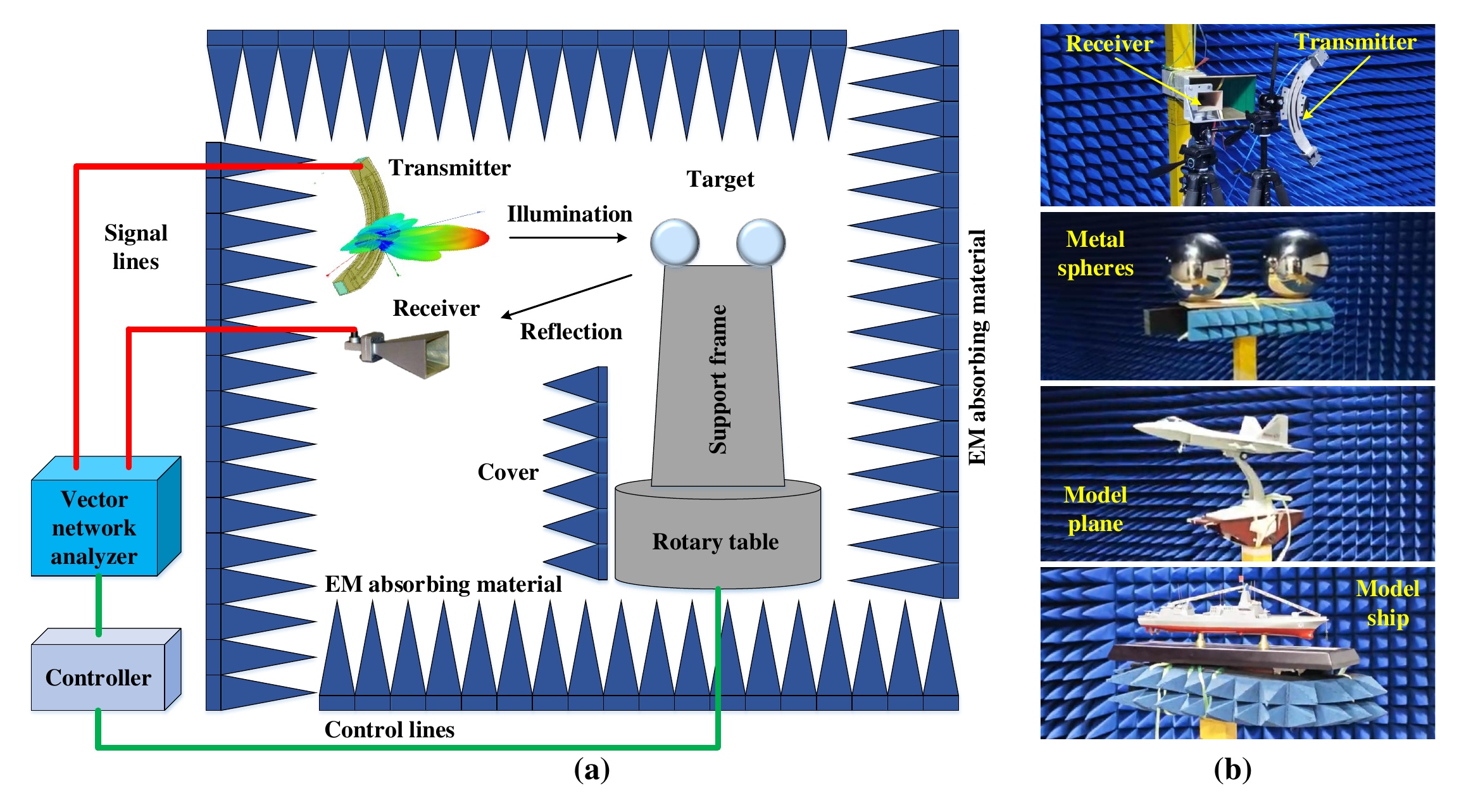}
\caption{RCS measurement scenarios. (a) Experiment setups in the microwave anechoic chamber. (b) Transceiver equipment and targets.}
\label{experiment_scenarios}
\end{figure}

As analyzed above, when an incident OAM wave is scattered by a target, the unique wavefront phase characteristics of the OAM wave cause its backscattered wave to differ from that of a plane wave. In other words, for a specific observation angle
$\phi $, the scattered plane wave may undergo destructive interference and cancel out in this direction, whereas the scattered OAM wave may experience constructive interference and amplify, resulting in a high echo peak. Notably, because different OAM modes have distinct phase wavefront distributions, the RCS of a target may vary across different OAM wave modes, leading to RCS diversity. In this study, we design an experiment to demonstrate the RCS differences between novel OAM waves without central energy void and traditional plane waves, as shown in Fig. \ref{experiment_scenarios}.
The transmitting and receiving antennas are both connected to two ports of a Vector Network Analyzer (VNA). The transmitting antenna can be a horn antenna for generating plane waves or the NTCS-OAM antenna proposed in this paper for generating OAM waves. A 10 GHz single carrier is used to illuminate the target. The receiving antenna is placed near the transmitting antenna, with the center observation angle defined at around ${90^ \circ }$. The target is mounted on a rotary table, completing a ${360^ \circ }$ step-wise rotation around its center at a constant speed. The rotary table and VNA are connected to a computer (controller), which coordinates the turntable rotation and VNA signal measurement to record the target's echo data from ${360^ \circ }$ observation angles. Then, the RCS of the target can be calculated based on the measurement parameters as
\begin{equation} \label{eq_experiment}
{\sigma _{{\rm{t}}\arg {\rm{et}}}}{\rm{ = }}\frac{{{P_{\rm{r}}} \cdot {{\left( {4\pi } \right)}^3} \cdot {R^4}}}{{{P_{\rm{t}}} \cdot {G_{\rm{t}}} \cdot {G_{\rm{r}}} \cdot {\lambda ^2}}},
\end{equation}
where, ${{P_{\rm{t}}}}$ is the Tx transmitting power, ${{P_{\rm{r}}}}$ is the Rx receiving power, ${{G_{\rm{t}}}}$ is the Tx antenna gain, ${{G_{\rm{r}}}}$ denotes the Rx antenna gain, $R$ indicates the range from the TX/Rx to the target, and $\lambda $ is the wavelength. The key parameters are shown in \ref{parameter}.
\begin{table}[!h]
\caption{Key experiment parameters.}
\begin{center}
\begin{tabular}{c|c|c}
\toprule
\textbf{Parameter} & \textbf{Value} & \textbf{Dimension} \\
 \midrule
 Central carrier frequency & 10.0 & GHz \\
 Transmission distance $R$ & 8.5 & m \\
 Target rotational speed & 1.0 & rad/min \\
 Tx transmitting antenna gain ${{G_{\rm{t}}}}$ & 16.0 & dB \\
 Rx receiving antenna gain ${{G_{\rm{r}}}}$ & 16.0 & dB \\
 Tx transmitting power ${{P_{\rm{t}}}}$ & 28.0 & dBm \\
 Rx receiving power ${{P_{\rm{r}}}}$ & 20.0 & dBm \\
 Same height of Tx/Rx and target & 1.5 & m \\
 Radius of the two metal spheres & 20.0 & cm \\
 Spacing between the metal spheres & 40.0 & cm \\
 \bottomrule
\end{tabular}
\end{center}
\label{parameter}
\end{table}

\begin{figure}[htbp]
\centering
\includegraphics[width=3.5in]{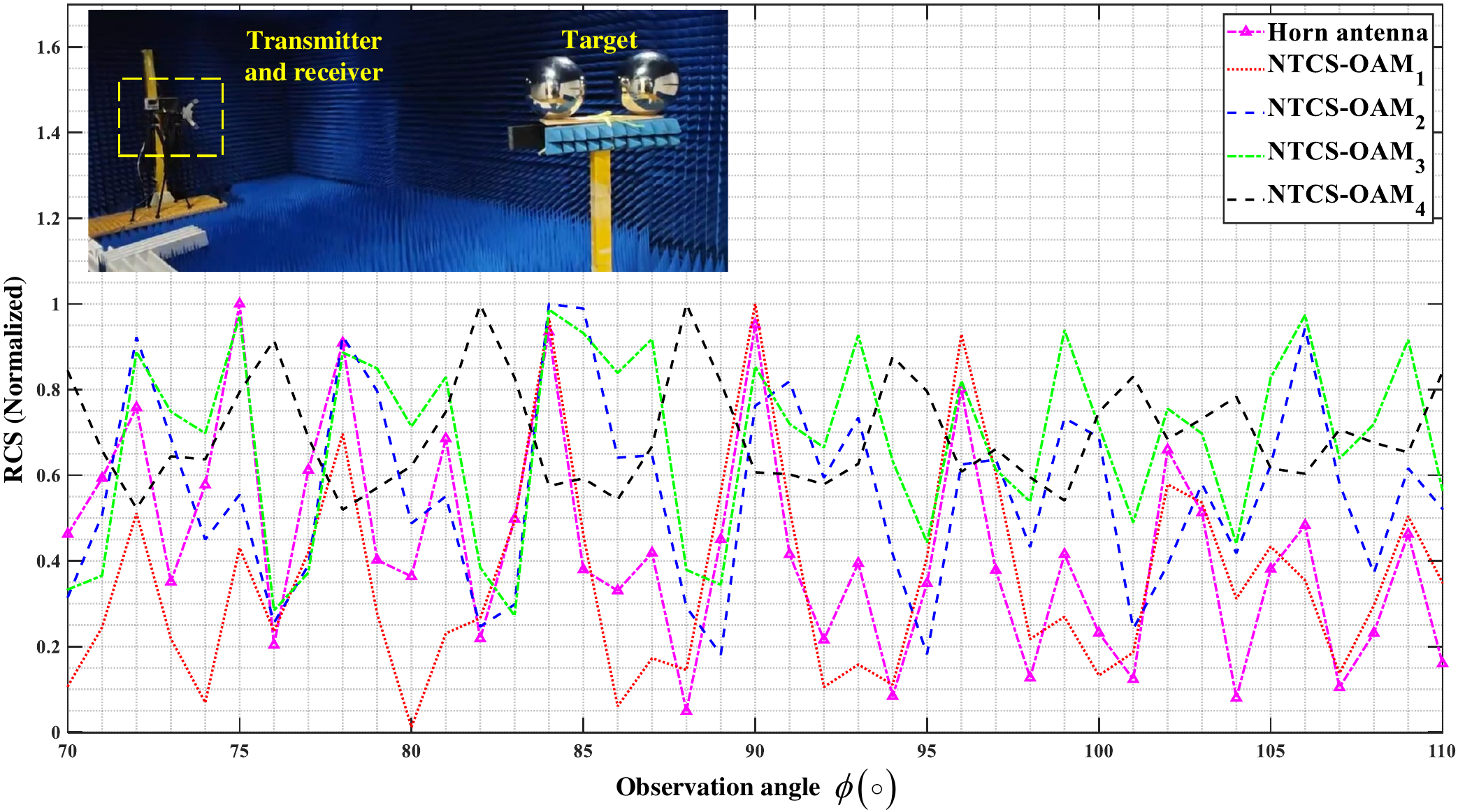}
\caption{Comparative experiments on the RCS diversity of the novel directional OAM beams, with the target being two identical metal spheres.}
\label{metal_spheres}
\end{figure}

The experiment results show significant angular dependence of RCS, with each OAM mode exhibiting unique patterns of fluctuation due to the spiral phase gradient of OAM EM waves. For instance, as illustrated in Fig. \ref{metal_spheres}, as the two metal spheres rotate around their center, the RCS demonstrates a fluctuating pattern at different observation angles. In the experiment, to avoid propagation path errors caused by antenna placement when switching between different transmitting antennas, we carefully aligned the echo peaks of different modes at a ${90^ \circ }$ observation angle. We then observed the echo fluctuations at other observation angles. The experimental results align with expectations, showing that the target exhibits distinct RCS diversity characteristics when illuminated by different OAM modes. This outcome can only be attributed to the varying wavefront phase gradients of the different OAM waves, eliminating the influence of placement errors.

\begin{figure}[htbp]
\centering
\includegraphics[width=3.5in]{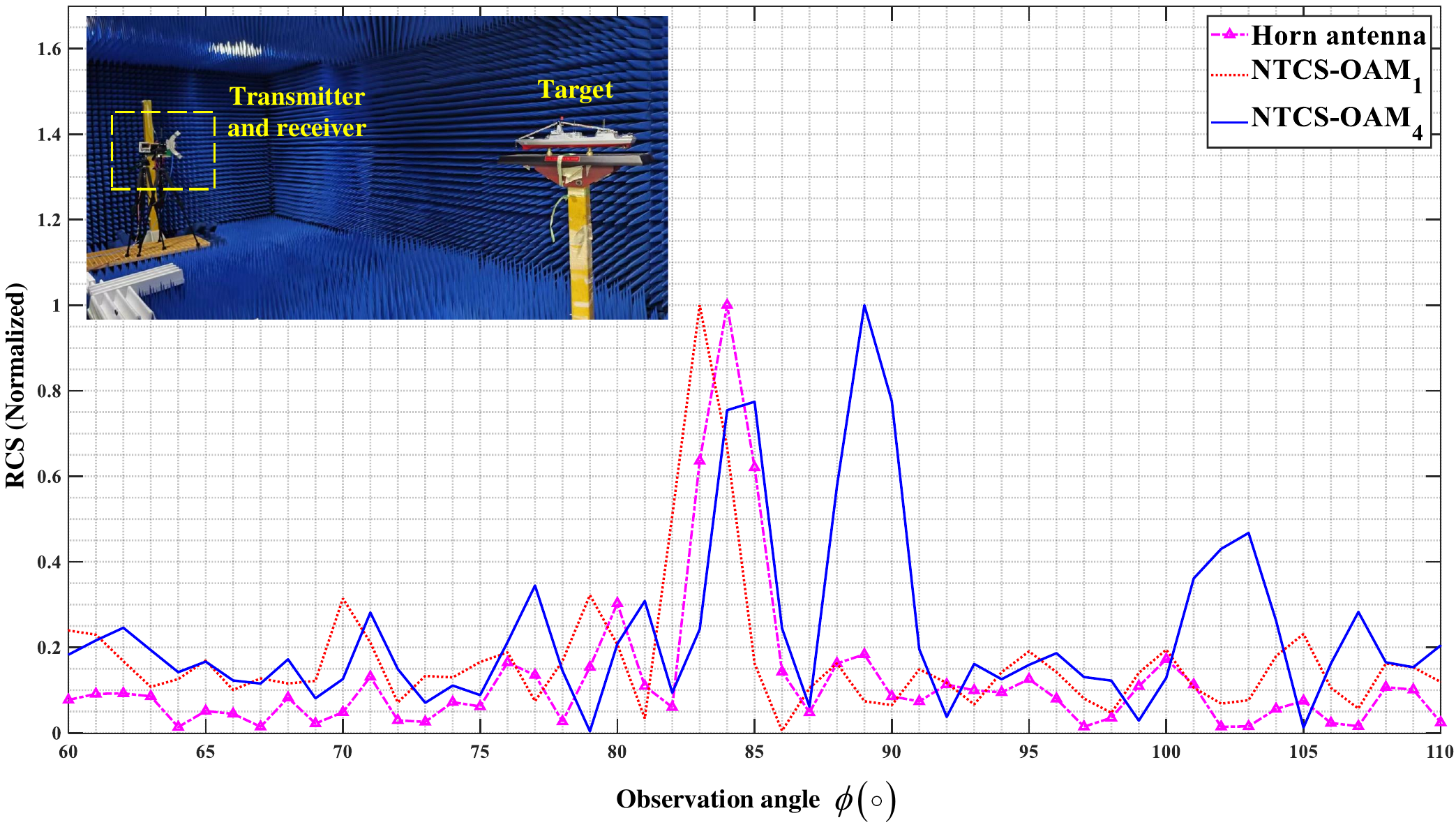}
\caption{Comparative experiment on the RCS diversity of the novel directional OAM beams, with the target being a model ship (made of metal).}
\label{model_ship}
\end{figure}

As shown in Fig. \ref{model_ship}, in addition to the simple metal spheres, we also conducted the same experiment using a model airplane. Because the model airplane has an asymmetrical structure, the RCS does not exhibit distinct periodic fluctuations as it rotates through ${360^ \circ }$ of observation angles. Instead, it displays significant echo peaks in the frontal observation direction. It can be seen that the angles at which peak echoes occur vary significantly under different OAM modes, further validating the RCS diversity phenomenon caused by the changes in various wavefront phase gradients.
\begin{figure}[htbp]
\centering
\includegraphics[width=3.5in]{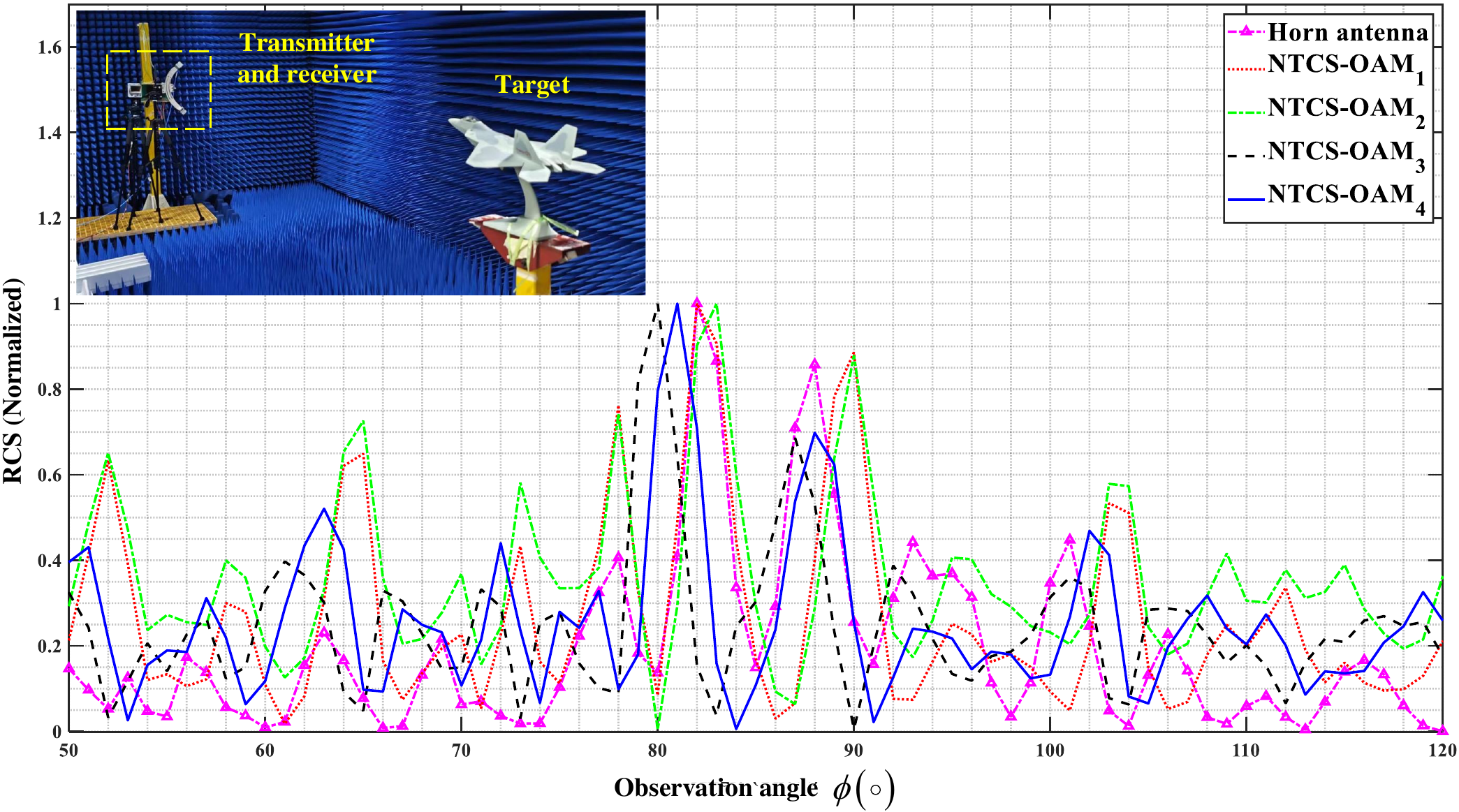}
\caption{Comparative experiment on the RCS diversity of the novel directional OAM beams, with the target being a model airplane (made of metal).}
\label{model_plane}
\end{figure}

To more clearly observe the differences in target echoes when illuminated by OAM waves versus plane waves, we selected only two OAM modes, $\ell  = 23$ and $\ell  = 45$, for the next experiment. The target was replaced with a metal model ship, and the experimental results are shown in Fig. \ref{model_ship}. As with the model airplane, it is evident that the echo peaks of the model ship are primarily concentrated in the frontal observation direction, where the contour is largest, and there are no longer any significant periodic RCS fluctuations at other positions. Compared to plane waves, the RCS results for the two different OAM modes exhibit significant angular differences. Specifically, at certain observation angles, plane waves do not produce peaks, whereas OAM waves can generate noticeable echo peaks. This indicates that OAM waves can be an effective tool for RCS measurements, providing additional degrees of freedom in the analysis and potentially enhancing target detection and characterization capabilities, resulting in additional diversity gains in target detection.

The experimental setup, especially the placement of transmitting and receiving antennas, can somehow affect results. To ensure accuracy, we carefully maintained the relative position of the OAM transmitting antenna and the receiving horn antenna when switching between different OAM modes. When measuring a structure made of two identical metal spheres, we aligned the echo peaks at a 90-degree observation angle and then analyzed the echoes from other angles. In practical monostatic radar systems, where a single antenna is used for both transmitting and receiving, these alignment issues do not exist. Our results demonstrate that OAM waves can create RCS diversity for the same target. However, the absolute values from the experiment are not highly significant, as echo strength can vary with different targets and distances.

In practical applications, the acquisition of RCS diversity information benefits from real-time dynamic wavefront manipulation, which places higher demands on reconfigurable antenna design. Reconfigurable intelligent surfaces (RIS) represent a highly promising solution. Previous studies have demonstrated that by dynamically adjusting the response state of each RIS unit in real-time, it is possible to achieve dynamic variations in OAM beams \cite{RIS,RIS2}. In the future, reconfigurable traveling-wave antenna designs could also be considered, with the goal of enabling dynamic adjustability of the phase variation slope within the main lobe of the wavefront. This potential remains an area for further in-depth investigation and exploration.

\section{Conclusion}
This study highlights the unique RCS diversity characteristics of the novel directional OAM waves compared to traditional plane waves, demonstrating their significant advantages in target detection. Through experiments with various targets, including metal spheres, model airplane, and model ship, we found that the novel directional OAM beams proposed in this paper exhibit distinct angular dependence due to their spiral phase gradient, leading to unique echo patterns and enhanced RCS diversity. Different OAM modes produce varying scattering characteristics, resulting in improved target detection and characterization capabilities. This suggests that OAM waves provide additional degrees of freedom for RCS measurements, overcoming limitations of conventional detection systems. The experimental findings underscore the potential of OAM to enhance wireless detection performance, providing a promising avenue for improved target identification in future intelligent transportation networks.


%





\ifCLASSOPTIONcaptionsoff
  \newpage
\fi

\end{document}